# Heavy Fermion-like metal α"-Fe$_{16}$N$_2$ with giant saturation magnetization

Nian Ji[1,2], Xiaoqi Liu[1,3], Jian-Ping Wang[1,2,3*]

[1] The Center for Micromagnetics and Information Technologies (MINT), University of Minnesota, Minneapolis, Minnesota, 55455, USA
[2] Department of Physics, University of Minnesota, Minneapolis, Minnesota, 55455, USA
[3] Department of Electrical Engineer and Computer Science, University of Minnesota, Minneapolis, Minnesota, 55455, USA

## Abstract

A new model is proposed for the strong ferromagnetism associated with partially localized orbitals in the Fe$_{16}$N$_2$ metallic system which draws substantially from models of heavy fermion metals. We demonstrated that an unusual correlation effect is brought up within the Fe-N octahedral cluster region and the effective on-site 3d-3d Coulomb interaction increases due to a substantial 3d electrons charge density difference between the clusters and its surroundings, which leads to a partially localized high spin electron configuration with a long range ferromagnetic order. First principle calculation based on LDA+U method shows that giant magnetic moment can be achieved at sufficiently large Hubbard U value. The feature of the coexistence of the localized and itinerant electron states plays a key role on the formation of the giant saturation magnetization.

[*] *Corresponding author Email: jpwang@umn.edu; Tel: 612-625-9509*

It is known that heavy fermion materials are always characterized by dual electronic states. They possess both localized orbitals which form the strong on-site magnetic moment and delocalized electrons which are responsible for the magnetic coupling and metallic properties. Traditional heavy fermion materials such as uranium containing compounds usually contain f electrons [1]. In some cases, transition metal oxides can also behave as heavy fermion metals. $LiV_2O_4$ [2] is reported to be a 3d electron heavy fermion system, and this challenges the conventional understanding. Consequently, it is logical to explore whether such electron behavior can occur in other materials. Indeed, this research is further justified by the s-d model [3], which involves a mixture of localized and itinerant electrons and is used to describe ferromagnetic materials in general.

The iron nitrides discussed here were discovered many years ago. Interest in their magnetic properties originates from a discovery by Kim and Takahashi in the 1970's [4], when they reported a giant saturation magnetization (Ms) observed on the phase α"-$Fe_{16}N_2$. Almost 20 years later, research on this topic was stimulated by Sugita et al. [5-7] based on single crystal phase α"- $Fe_{16}N_2$ thin films grown epitaxially on GaAs or In doped GaAs substrates by Molecule Beam Evaporation (MBE). Saturation magnetization is found to be 3.2T, with a corresponding magnetic moment as high as $3.5\mu_B$/Fe at low temperature. However, the subsequent investigations from many other groups on bulk and thin films samples containing α"- $Fe_{16}N_2$ led to conflicting results with Ms value less than 2.3T (2.5 $\mu_B$/Fe) [8-11]. Somewhat higher moments have been reported in films prepared by specially designed sputtering beams or facing target sputtering systems [12,13], reaching a value up to $2.9\mu_B$/Fe and $3.0\mu_B$/Fe, respectively. Due to such inconsistent and poorly reproducible reports, $Fe_{16}N_2$ has been regarded a mystery among magnetic researchers.

Initial theoretical models treated α"- $Fe_{16}N_2$ as a metal with itinerant magnetism, and so used local spin density approximation (LSDA) based electronic structure calculations to predict its magnetic properties. These calculations do not predict high magnetic moments in $Fe_{16}N_2$ even when first principle methods [14] or nonlocal correction have been considered [15]. When nitrogen is added to iron, in the LSDA scheme, the down spin electron changes distribution on different iron sites due to p-d hybridization between N and its nearest 6 iron neighbors [16]. Measurements on other iron nitrides phases suggest quantitative agreement with this down spin electron redistribution process. However, it can not increase the average magnetic moment above 2.7 $\mu_B$/Fe for the α" phase. To rationalize the high moment reports, Lai et al. [17] postulated the existence of a strong correlation effect on all iron sites in $Fe_{16}N_2$ system and obtained the magnetic moment more closely resembling the high moment experimental data. However, people started to throw doubts on whether the material system can maintain its metallic properties given the high on-site Coulomb U values and its inability to guide the experiment. A. Sakuma proposed a high moment scenario called the charge transfer model [18], inferring the existence of empty nitrogen orbitals near the Fermi level, which serve as charge hopping sites. This qualitative model predicts a high spin configuration of Fe and long range ferromagnetic order through an effective "double exchange" process in the context of highly localized spin interacting picture. However, no obvious evidence shows the N site is much more negatively charged relative to other iron nitrides. Recently, our group has performed X-ray magnetic circular dichroism (XMCD) measurements on high-moment $Fe_{16}N_2$ samples and discovered the localized 3d electron states 19], which provide further insights and experimentally confirm the α"- $Fe_{16}N_2$ system to have a dual electron configuration much like the heavy fermion system. This observation implies that neither itinerant nor localized magnetism can satisfyingly describe the mechanism in $Fe_{16}N_2$ system. In this paper, a new model is provided to address the longstanding giant saturation magnetization problem based on the idea of 3d electron partial localization. It is shown that there appears a considerable charge density difference inside and outside the Fe-N clusters due to the reduced symmetry in the $Fe_{16}N_2$ system. As a result, the electron correlations interior and exterior the cluster region possess different nature. Local Spin Density Approximation with on-site Coulomb (U) correction (LDA+U) based calculation is performed to illustrate the essential features of the physical picture discussed here. As expected, the origin of giant saturation magnetization and strong ferromagnetism associated with high spin configuration can be realized for sufficiently large U values.

The crystal structure of $Fe_{16}N_2$ is extensively discussed by K. H. Jack [20], which contains both bct-Fe and fcc-$Fe_4N$ like local Fe environment. In both bct-Fe and fcc $Fe_4N$, a D4h symmetry Fe is formed due to the lattice expansion along [001] direction. However, in a real $Fe_{16}N_2$ system, the Fe sites with local D4h point group symmetry reduce to C4v and C1h symmetry due to the alternating occupancy of the N atoms. Unlike $Fe_4N$ or bct Fe, the corner Fe sites within those distorted Fe-N octahedrons lose the mirror reflection symmetry along certain directions, experiencing less symmetric crystal field due to the center N. Therefore, an additional effect arises due to the symmetry break. Plotted in fig.1 is the total charge density for $Fe_4N$ and $Fe_{16}N_2$ on the Fe-N plane perpendicular to the [001] direction based on LDA method. As illustrated in the figure, in $Fe_4N$, Fe-N clusters are connected through the corner Fe, which results in relative uniform charge distribution along line CD. In the contrary, in $Fe_{16}N_2$, Fe-N clusters are geometrically isolated in the sense that each corner Fe site belongs to an individual Fe-N cluster. As a result, there appears a significant charge density difference interior and exterior the cluster region which originates from this reduced symmetry. These observations lead us to suggest 3d electron partial localization model for the strong ferromagnetism in this $Fe_{16}N_2$ material system.

The starting point for a discussion of the partial localization model is that there exists a considerable 3d electron charge density difference between iron sites inside and outside the cluster region. To illustrate the charge density effect on its magnetic properties, a model calculation is performed to capture the essential qualitative features. Considering a uniform d electron charge distribution within the cluster as illustrated schematically in fig.2a, the static potential difference between interior and exterior of the clusters can be obtained by solving Poison's equation. This potential can be written as $\frac{2\pi}{3}(\rho_{in} - \rho_{out})a_0 r_0^2 Ry$. With $\rho_{in}$ and $\rho_{out}$ represent the d electron charge density interior and exterior the sphere respectively, $a_0$ the Bohr radius and $Ry$ as the Rydberg constant. This potential energy difference is created solely by the short-region repulsive electron interaction as a screened Coulomb energy in electron gases, which qualifies an on-site Coulomb interaction difference between the cluster region and its metallic environment (Fe4d). The U values found in metallic Fe are typically 1ev [21]. Plotted in fig.2b is the effective U value within the cluster versus $n$, which is defined as the average number of 3d electron /Fe of the iron site inside the cluster subtract that outside the cluster. According to our model, U increases linearly as $n$ increases, reaching a value of 7~8ev at $n$~ 0.5. In the LSDA picture, the number of 3d electrons/Fe on Fe4d, Fe4e and Fe8h sites calculated by Jingsong He *et al* [22] is 6.65, 6.57 and 6.58, respectively. In our model, if ~0.4 3d electrons transfer from Fe4d site to its four neighboring Fe-N octahedral clusters, it will give rise to a number of 3d electrons /Fe of 6.25, 6.70 and 6.71 for Fe4d, Fe4e and Fe8h respectively and $n$~0.5 can be easily reached. At this point, the physical picture is rather clear. Certain amount of 3d electrons transfers from Fe4d site to Fe-N octahedral cluster and creates a local potential energy fluctuation, which results in a large U value inside the cluster relative to its outside region. To stabilize this charge transfer process, additional energy term is required to compensate the U increase. Therefore, the bonding between Fe and N plays an important role within the cluster. In the atomic limit, the total energy of the N atom will be lowered if its empty p orbitals are filled by additional electrons. In the Fe-N cluster, similar scenario may be realized in the sense that iron and nitrogen form hybridized orbitals to accommodate electrons from Fe4d sites, which offer total energy lower enough to overcome the U value increase. In analogy to the Stoner's ferromagnetism stability argument, such a high

U value leads to the reduction of the band overlapping between spin up and down and effectively increases the exchange energy. Ultimately, certain orbitals within the Fe-N cluster will possess localized structure with fully polarized spin configuration, where Hund's coupling dominates. On the other hand, the other orbitals remain broad band width and less exchange splitting due to their spatial extension and strong hybridization with neighboring metallic Fe4d sites outside the sphere. This condition helps to maintain their band-like features that favor long-range ferromagnetic coupling and metallic behavior. In that sense, the dual electron behavior and partial localization is realized. To illustrate this idea quantitatively further, an electronic structure calculation with on-site Coulomb interaction corrections is performed to compute magnetic moment in $Fe_{16}N_2$. To simulate partial localization, we used the LDA+U method used by Anisimov et al [23]. The U value of Fe outside (Fe4d) the cluster sphere was fixed at 1 ev and the magnetic moment change was then plotted for three Fe sites as the U inside the cluster (same U is assumed for both Fe4e and Fe8h) varied as shown in fig.3a. The J parameter (Stoner parameter) was taken to be U/10. Calculation is performed by using the Vienna *ab initio* simulation package (VASP), with Projector Augmented Wave potential [24]. A plane wave cutoff corresponds to kinetic energy of 250ev was used. A k mesh of 8*8*8 was used to sample the supercell. The exchange correlation functional is approximated with the fully localized limit of the local spin density approximation +U method [25]. The lattice constant is constrained according to its bulk value. The density of states (DOS) were calculated with fine k meshes using tetrahedron method with Blöchl corrections as implemented in VASP. Both Fe4e and Fe8h sites are sitting inside the octahedral cluster region. As the U increases, they favor rapid polarization while for the Fe4d site, which sites outside the Fe-N cluster, exhibits slow decay of the moment from 2.9 to 2.6$\mu_B$/Fe and levels off at high U values. An average magnetic moment of ~3$\mu_B$/Fe can be obtained if U=8ev is assumed. The effect of the d-d coulomb interaction on the electronic structure can be explained in the following. Plotted in fig. 3b is the partial DOS projected on double degenerate (dxz+dyz) orbital of Fe4e sites. As U begins to increase, the spin up band shifts downwards with more and more states occupied at the expense of spin down orbital in the process of self-consistency iteration. This orbital will eventually become fully occupied when U is sufficiently large. In that case, electrons on these orbitals exhibit nearly atomic configuration and the on-site magnetic coupling mechanism is dominated by Hund's rule, which gives rise to a high spin configuration of Fe. It is also interesting to notice that the Fe4d sites reduce their magnetic moment as U increases. This suggests that the electron distribution on different iron sites changes significantly as opposed to the LSDA calculation results, in which case, the high moment on Fe4d sites are due to the charge transfer from Fe4d to the Fe sites closer to the N sites.

The essential difference between our model and others' is the idea of 3d electron partial localization. The LDA+U methods are initially developed to treat Mott insulator systems such as FeO. The earlier Lai's model considers a strong correlation effect on all three iron sites and achieves high magnetic moment. The U values used are calculated from the embed cluster method [24] by assuming a relatively small screening constant, which has been criticized as "unrealistic" for a metal system [26]. Experimentally, low temperature transport properties suggest the legitimacy of treating $Fe_{16}N_2$ as a metal [27]. However, large residual resistivity is also explored relative to the pure metal Fe, implying the existence of defects-like localized electron states though single crystal samples are used for this study. Recently, our group has performed similar measurements that confirmed this observation [28]. In our model, we emphasize that the unusual correlation occurs within the Fe-N octahedrons due to the introduction of the center N atom. The metallic behavior of the system still stands as the U value on the Fe4d sites remain metal like. More importantly, the difference of the on-site Coulomb U for the three iron sites distinguishes the material into two subsets, namely, Fe-N octahedrons (Fe8h and Fe4e) and metallic Fe atoms (Fe4d). Within the clusters, Fe8h and Fe4e possess a partially localized electron behavior due to the enhancement on the U value. Outside the cluster region, Fe4d maintain its metal like behavior. The macroscopic giant ferromagnetic order is formed through the interplay between localized oribitals and itinerant band. Therefore, to rationalize the giant saturation magnetization report, a dual electron behavior seems to be a necessity in a fundamental level. This opens up the direction on identifying promising candidates for many applications that demand high $M_s$.

In conclusion, calculations using LDA+U method gives a theoretical justification for the proposed solution on the long standing $Fe_{16}N_2$ problem and additionally offers an explanation of the experimental discovery on the coexistence of the itinerant and localized 3d electrons in $Fe_{16}N_2$ system. The partial localization concept provides insight to the effect brought about by N sites. In particular, isolated Fe-N octahedral clusters induce non-uniform charge distribution and cause an increase of effective U value that generates both localized and itinerant d electrons. This feature allows the system to have global ferromagnetic coupling while locally maintain Hund's rule dominated magnetic moments.

The work was partially supported by the U.S. Department of Energy, Office of Basic Energy Sciences under contract No. DE-AC02-98CH10886, National Science Foundation NNIN program at University of Minnesota. The authors would like to thank Prof. Jack Judy, Dr. Mark Kief and Dr. Yinjian Chen for the useful discussion. We are also grateful for resources from the University of Minnesota Supercomputing Institute.

**Figure Captions**

FIG.1. Total charge density distribution projected on Fe-N plane perpendicular to the [001] plane for $Fe_{16}N_2$ and $Fe_4N$ based on LDA calculation results, the right figures shows intensity of charge density (logarithmic scale and normalized to their maximum intensities) along line AB and CD as marked in the charge figure is shown to illustrate the effect of the broken symmetry as explained in text.

FIG.2a The crystal structure of $Fe_{16}N_2$. Fe-N octahedral clusters are highlighted in red color. The proposed charge density distribution is schematically marked by yellow sphere. $\rho_{in}$ and $\rho_{out}$ as marked on the figure is to illustrate such difference. The right figure shows chemical environment of the three iron sites, Fe8h and Fe4e are within the cluster region and Fe4d locates outside the cluster.

FIG.2b 3d electron charge difference per iron site between interior and exterior the Fe-N cluster sphere versus the effective on site Coulomb energy within the cluster.

FIG. 3 a) Magnetic moment of different Fe site in $Fe_{16}N_2$ vs U as explained in text. b) Partial density of states projected on (dxz+dyz) of Fe4e site in $Fe_{16}N_2$ for different U values.

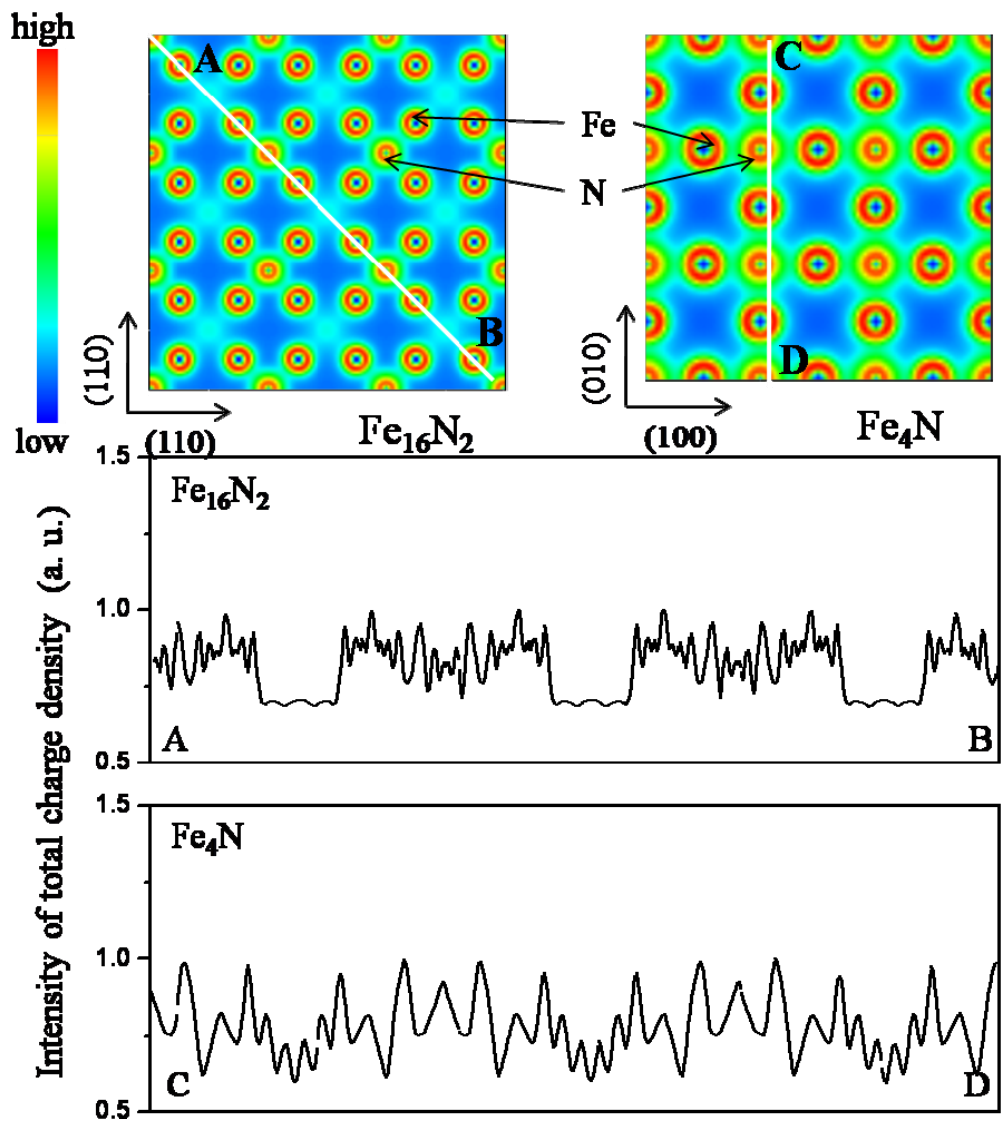

FIGURE 1 Total charge density distribution projected on Fe-N plane perpendicular to the [001] plane for Fe$_{16}$N$_2$ and Fe$_4$N based on LDA calculation results, the right figures shows intensity of charge density

(logarithmic scale and normalized to their maximum intensities) along line AB and CD as marked in the charge figure is shown to illustrate the effect of the broken symmetry as explained in text.

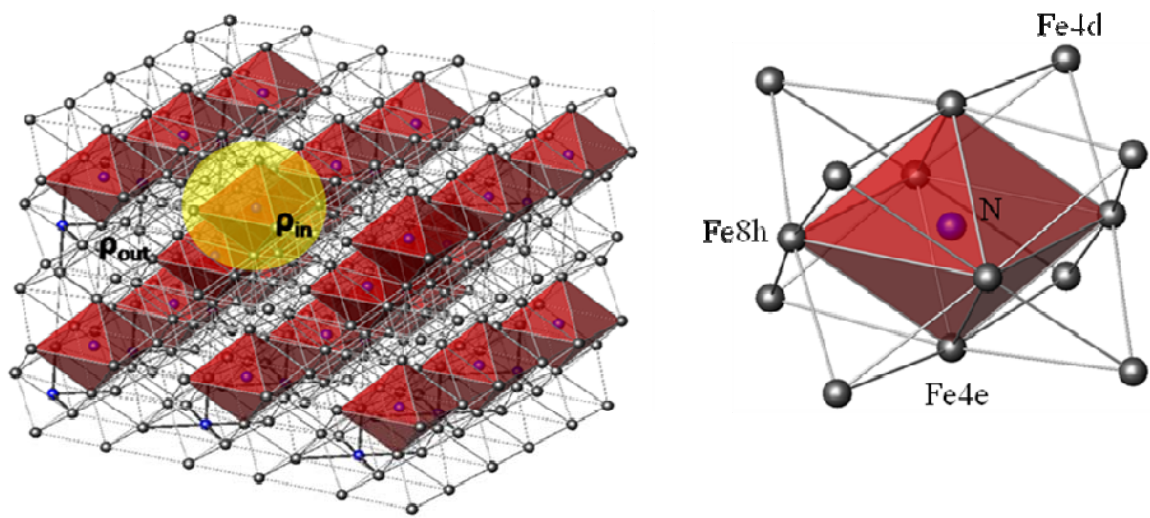

FIGURE 2a. The crystal structure of $Fe_{16}N_2$. Fe-N octahedral clusters are highlighted in red color. The proposed charge density distribution is schematically marked by yellow sphere. $\rho_{in}$ and $\rho_{out}$ as marked on the figure is to illustrate such difference. The right figure shows chemical environment of the three iron sites, Fe8h and Fe4e are within the cluster region and Fe4d locates outside the cluster.

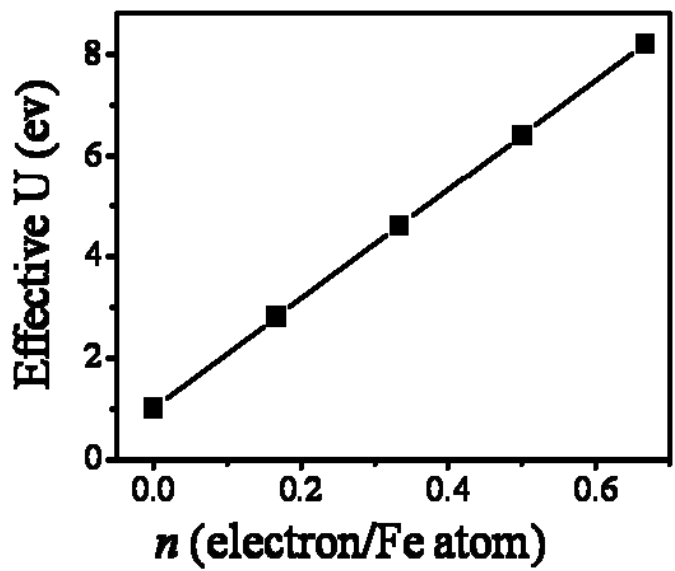

FIGURE 2b 3d electron charge difference per iron site between interior and exterior the Fe-N cluster sphere versus the effective on site Coulomb energy within the cluster.

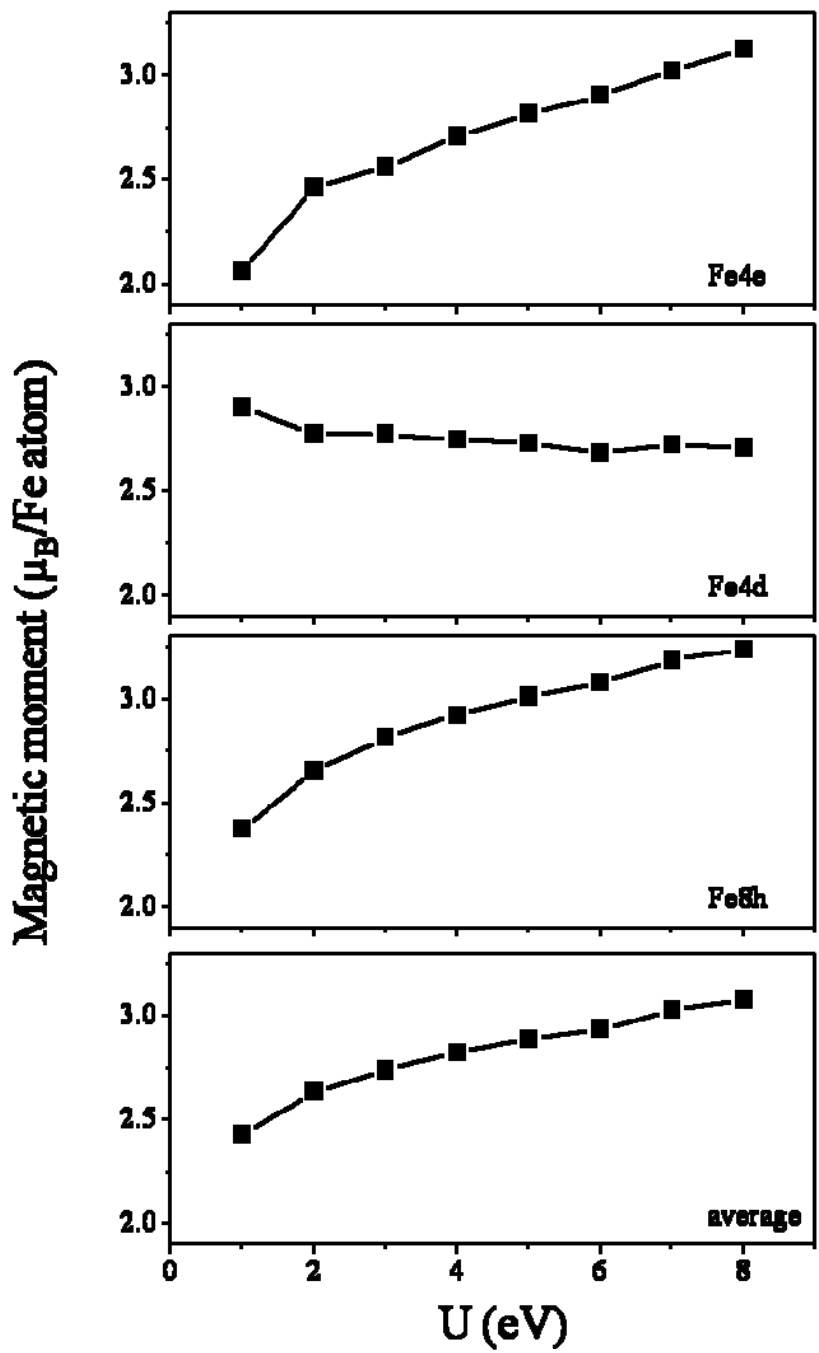

FIGURE 3a Magnetic moment of different Fe site in $Fe_{16}N_2$ vs U as explained in text.

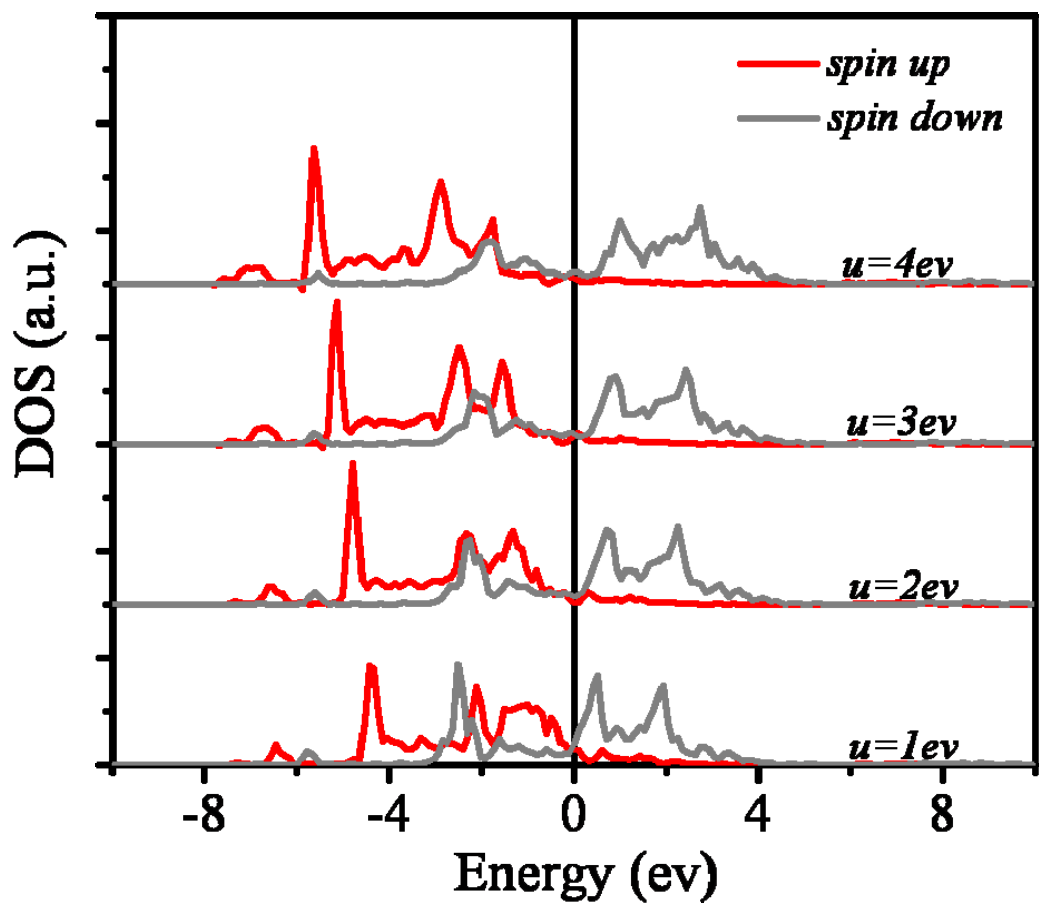

FIGURE.3b Partial density of states projected on (3dxz+3dyz)/2 of Fe4e site in $Fe_{16}N_2$ for different U values.